# Stochastic diffusion of electrons interacting with whistler-mode waves in the solar wind

Tien Vo,  Robert Lysak and  Cynthia Cattell

View Online    Export Citation    CrossMark

## ARTICLES YOU MAY BE INTERESTED IN

# Stochastic diffusion of electrons interacting with whistler-mode waves in the solar wind

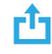
View Online　　Export Citation　　CrossMark

Tien Vo,[a] (ID) Robert Lysak, (ID) and Cynthia Cattell (ID)

## AFFILIATIONS

School of Physics & Astronomy, University of Minnesota, Minneapolis, Minnesota 55455, USA

[a]Author to whom correspondence should be addressed: Tien.Vo@colorado.edu

## ABSTRACT

The effects of increasing the whistler amplitude and propagation angle are studied through a variational test particle simulation and calculations of the resonance width. While high amplitude and oblique whistlers in typical 1 AU solar wind parameters are capable of forming an isotropic population without any additional processes, anomalous interactions with quasi-parallel whistlers may be essential to the process of halo formation near the Sun. High amplitude and quasi-parallel whistlers can scatter strahl electrons to low velocities (less than the wave phase velocity) to form a halo population, as long as their amplitude is sufficiently high. We also present in detail a careful treatment of the sensitivity to initial conditions based on calculations of the phase space volume, which is necessary for numerical calculations of highly stochastic motion due to resonant interactions with large amplitude waves. Our method ensures that the volume-preserving characteristic of the Boris algorithm is consistently applied for simulations of both stochastic and non-stochastic particle motion.

*Published under an exclusive license by AIP Publishing.* https://doi.org/10.1063/5.0074474

## I. INTRODUCTION

It has been a long-standing interest to identify the mechanisms that regulate the solar wind heat flux, mainly carried by electrons. Solar wind electrons typically contain three populations: a thermal, isotropic core; a suprathermal, isotropic halo; and a suprathermal, anisotropic tail formed by field-aligned "strahl" electrons streaming anti-sunward (Montgomery et al., 1968; Feldman et al., 1975). While the core represents the bulk of the electrons, the suprathermal populations carry most of the heat flux into interplanetary space (Pilipp et al., 1987; Halekas et al., 2021).

Observations show that the halo is almost nonexistent in the young solar wind (Halekas et al., 2020), and its relative density (with respect to the local total density) increases radially in anticorrelation with the strahl relative density (Maksimovic et al., 2005; Štverák et al., 2009) from near the Sun (≤0.3 AU) to beyond 1 AU, resulting in either a highly broadened or nonexistent "strahl" electrons at large distances (Anderson et al., 2012; Graham et al., 2017; Berčič et al., 2019; 2020). These observations suggest that there are some mechanisms at play to counter adiabatic focusing effects, which would otherwise lead to an opposite radial trend.

The contemporary agreement is that the halo formation and strahl depletion/broadening are correlated (Halekas et al., 2021; Cattell et al., 2021a). Various scattering mechanisms may play a role in regulating the heat flux (López et al., 2020). The major candidates are collisionless heat flux instabilities (HFIs) involving electromagnetic whistler-mode waves. The whistler HFI, which is the fastest growing mode compared to other HFIs in typical solar wind conditions (Gary et al., 1994; Gary et al., 1999), generates quasi-parallel whistlers through cyclotron resonance at the velocity range of the halo (Verscharen et al., 2019; Tong et al., 2019b). In the quiet solar wind at 1 AU, these waves are often observed with small ($\delta B/B_0 \lesssim 0.01$) amplitudes (Lacombe et al., 2014; Tong et al., 2019a) and are mainly effective in scattering electrons outside of the strahl velocity range. Therefore, theoretical arguments and simulations cast doubts on their ability to scatter strahl electrons (López et al., 2019; Verscharen et al., 2019). For this reason, interest has shifted to whistlers that propagate obliquely, which are observed by satellites both near and far from the Sun.

High amplitude ($\delta B/B_0 \gtrsim 0.1$), oblique whistlers were first observed near stream interaction regions (SIRs) at 1 AU with (maximum) electric fields greater than 40 mV m$^{-1}$ with an average of ∼10 mV m$^{-1}$ (Breneman et al., 2010; Cattell et al., 2020). At less than 0.3 AU, these waves have also been seen at the same range of amplitudes (tens of mV m$^{-1}$) together with evidence of strahl pitch angle width broadening (Cattell et al., 2021b). While quasi-parallel whistlers only have small amplitudes at 1 AU, large amplitude whistlers, both quasi-parallel and oblique, have been reported near the Sun (Agapitov et al., 2020; Cattell et al., 2021b). The oblique propagation angle enables anomalous resonant interactions at the strahl velocity range, accelerating field-aligned electrons to larger pitch angles (Vasko *et al.*, 2019; Verscharen *et al.*, 2019). The high amplitude leads to resonance overlaps and allows electrons to diffuse stochastically through a wide range of pitch angles (Karimabadi *et al.*, 1990; Karimabadi *et al.*, 1992).

The effective scattering of field-aligned electrons by high amplitude, oblique whistlers has been demonstrated through self-consistent particle-in-cell (PIC) simulations with solar flare parameters (Roberg-Clark *et al.*, 2016; 2018; 2019). Micera *et al.* (2020) simulated these whistlers in the pristine solar wind at 0.3 AU and showed that the formation of an isotropic halo from strahl electrons was possible. In their simulation, while oblique whistlers were initially generated and scattered strahl electrons, quasi-parallel whistlers appeared at later stages to fully isotropize the velocity distribution function (VDF). However, details of this two-stage scattering process were not fully described. While quasi-parallel whistlers were necessary at the last stage to form a fully isotropic halo, the velocity range in which they were effective was not determined.

Cattell and Vo (2021) performed test particle simulations with oblique whistlers in multiple solar wind parameters, thereby studying only the first stage of the process of halo formation (strahl scattering). Simulations with parallel whistlers and more realistic wave profiles (packets of frequencies) were also shown. In their simulations with 0.3 AU parameters (consistent with those in Micera's simulation), there was a limit to the strahl scattering due to oblique, both monochromatic and narrowband, whistlers. This suggests that the distribution will be fully isotropized only if the amplitude of quasi-parallel whistlers is large enough so that their effective velocity range overlaps with that of oblique whistlers. In the later stages of Micera's simulation, quasi-parallel whistlers had $\delta B/B_0 \gtrsim 0.1$ (Micera, private communication), providing grounds for this hypothesis.

In this paper, we use test particle simulations to show explicitly that high amplitude, quasi-parallel whistlers have a large effective velocity range, utilizing calculations of the resonance width. We demonstrate the effects of increasing whistler amplitude and propagation angle and show that high amplitude, quasi-parallel whistlers are not essential to the isotropization of the VDF at 1 AU as they are near the Sun. We also present our numerical methods, which were only briefly outlined in Cattell and Vo (2021). In Sec. II, we discuss the relevant theory of wave–particle interaction and the stochasticity (sensitivity to initial conditions) in simulations of our system. In Sec. III, we describe our numerical methods and present a careful treatment of stochastic particle solutions. In Sec. IV, we show our simulation results and compare the resonance width with the analytical prediction. In Sec. V, we discuss the physical implications of our results and provide concluding remarks.

## II. THEORY

### A. Hamiltonian formulation of resonant interaction

Karimabadi *et al.* (1990), hereby referred to as K1990, provided a general treatment of wave–particle resonant interactions using the secular perturbation theory, some results of which are relevant for later physical discussions and will be quoted here. In a cold uniform plasma, a monochromatic whistler-mode wave has an electromagnetic field with

$$\mathbf{B}_w = B_x^w \sin\psi \hat{\mathbf{x}} + B_y^w \cos\psi \hat{\mathbf{y}} + B_z^w \sin\psi \hat{\mathbf{z}}, \quad (1a)$$

$$\mathbf{E}_w = E_x^w \cos\psi \hat{\mathbf{x}} - E_y^w \sin\psi \hat{\mathbf{y}} + E_z^w \cos\psi \hat{\mathbf{z}}, \quad (1b)$$

where $\psi = \mathbf{k} \cdot \mathbf{r} - \omega t = k_\perp x + k_\parallel z - \omega t$ is the wave phase, $k_\perp = k \sin\alpha$, and $k_\parallel = k \cos\alpha$. The polarizations, derived from the cold plasma dispersion relation, are given in Tao and Bortnik (2010). These fields correspond to a scalar potential $\Phi_w = \Phi_0 \sin\psi$ and a vector potential $\mathbf{A}_w = A_1(k_\parallel/k) \sin\psi \hat{\mathbf{x}} + A_2 \cos\psi \hat{\mathbf{y}} - A_1(k_\perp/k) \sin\psi \hat{\mathbf{z}}$, where

$$\Phi_0 = -\frac{1}{k}\left[\left(\frac{k_\perp}{k}\right)E_x^w + \left(\frac{k_\parallel}{k}\right)E_z^w\right], \quad (2a)$$

$$A_1 = \frac{1}{\omega}\left[\left(\frac{k_\parallel}{k}\right)E_x^w - \left(\frac{k_\perp}{k}\right)E_z^w\right], \quad (2b)$$

$$A_2 = \frac{E_y^w}{\omega}, \quad (2c)$$

such that $\mathbf{E}_w = -\nabla\Phi_w - \partial\mathbf{A}_w/\partial t$. Assume also a uniform background field $\mathbf{B}_0 = B_0\hat{\mathbf{z}}$. Then, the relativistic Hamiltonian of an electron with charge $e$ and mass $m$ is $\mathcal{H} = \sqrt{m^2 c^4 + (\mathbf{P} + e\mathbf{A}_w + m\Omega_c x\hat{\mathbf{y}})^2 c^2} - e\Phi_w$ where $\mathbf{P} = \gamma m\mathbf{v} - e\mathbf{A}_w - m\Omega_c x\hat{\mathbf{y}}$ is the canonical momentum conjugate to the Cartesian coordinates, $\Omega_c = eB_0/m$ is the cyclotron frequency, and $\gamma = (1 - v^2/c^2)^{-1/2}$ is the Lorentz factor.

Let the normalized wave amplitudes $\varepsilon_{1,2} = eA_{1,2}/mc$ and $\varepsilon_3 = e\Phi_0/mc^2$ be small ($|\varepsilon_{1,2,3}| \ll 1$). Through two consecutive canonical transformations, first into the guiding center frame and second into the rotating wave frame (details in the Appendix of K1990), the gyroaveraged Hamiltonian can be written up to first order in $\varepsilon$ as $\mathcal{H} = \mathcal{H}_0 + \mathcal{H}_1$, where

$$\mathcal{H}_0 = \gamma mc^2 - (\omega/k_\parallel)\hat{P}_\parallel, \quad (3)$$

and the perturbation $\mathcal{H}_1 = Z_n \cos(k_\parallel\hat{z})$ has an amplitude $Z_n$ in terms of the $n$th order Bessel function of the first kind $J_n$ and its derivative $J_n'$ as follows:

$$Z_n = \frac{mc^2}{\gamma}\left[\varepsilon_1\left(-\frac{\hat{P}_\parallel}{mc}\sin\alpha + \frac{n\Omega_c}{ck_\perp}\cos\alpha\right)J_n(k_\perp\hat\rho)\right.$$
$$\left. + \varepsilon_2\sqrt{\frac{\hat{P}_\perp^2}{m^2 c^2} + \frac{2n\Omega_c}{ck_\parallel}}J_n'(k_\perp\hat\rho) - \gamma\varepsilon_3 J_n(k_\perp\hat\rho)\right]. \quad (4)$$

$n \in \mathbb{Z}$ is the harmonic of the cyclotron frequency (primary resonance of the interaction). The transformed coordinates are $\hat{P}_\parallel = P_\parallel = P_z$ and $\hat{z} = z - (\omega/k_\parallel)t - (k_\perp/k_\parallel)(P_y/m\Omega_c) + n\theta/k_\parallel - \pi/2k_\parallel$ where $\theta$ is the gyrophase. The gyroradius $\hat\rho$ and $\hat{P}_\perp$ are exactly defined as in K1990. Note that the results here are written in SI instead of cgs units.

To the zeroth order, $\mathcal{H} \approx \mathcal{H}_0$ is invariant. Thus, the electron motion is mostly constrained on a constant energy (H) surface defined by (3). Resonant interactions come from first order effects (the perturbation $\mathcal{H}_1$). Examining the equation of motion around the fixed points $(\hat{z}_r, \hat{P}_{\parallel|r})$ satisfying $d\hat{z}/dt = 0$ and $d\hat{P}_\parallel/dt = 0$ leads to the following resonance condition:

$$\omega - k_\parallel\hat{P}_\parallel/m\gamma - n\Omega_c/\gamma = 0, \quad (5)$$

which defines the momentum $\hat{P}_{\parallel r} \approx \gamma m v_{\parallel r}$ for a given harmonic $n$ at which the electrons interact resonantly. Expanding $\mathscr{H}$ around these points yields the Hamiltonian of a pendulum with torque (see Sec. IV of K1990). Thus, resonant particles oscillate quasi-periodically on an H surface corresponding to their initial conditions. The amplitude of such oscillations is called the trapping half width (or resonance width), defined by

$$
\begin{aligned}
\Delta \hat{P}_{\parallel r} &= 2mc \left| \frac{Z_n}{\partial^2 \mathscr{H}_0 / \partial \hat{P}_{\parallel}^2} \right|^{1/2} \\
&= \frac{2mcN_{\parallel}}{|N_{\parallel}^2 - 1|^{1/2}} \left| - \left( \frac{P_{\parallel}}{mc} \varepsilon_1 \sin \alpha + \gamma \varepsilon_3 \right) J_n \right. \\
&\quad + \frac{1}{2} \frac{P_{\perp}}{mc} [(\varepsilon_2 + \varepsilon_1 \cos \alpha) J_{n-1} - (\varepsilon_2 - \varepsilon_1 \cos \alpha) J_{n+1}] \Big|^{1/2} , \quad (6)
\end{aligned}
$$

where $N_{\parallel} = ck_{\parallel}/\omega$ and $P_{\parallel}, P_{\perp}, \gamma, \hat{\rho}$ are evaluated near a resonance defined by Eq. (5).

The transition from quasi-periodic (regular) to stochastic motion occurs for large wave amplitudes. When the widths $\Delta \hat{P}_{\parallel r}$ of two adjacent resonances overlap, the particles are no longer trapped and can diffuse stochastically across multiple harmonics. The separation on an H surface between two consecutive resonances is $\delta \hat{P}_{\parallel} = mc(\Omega_c/\omega)[N_{\parallel}/(1 - N_{\parallel}^2)]$. Thus,

$$
C = (2\Delta \hat{P}_{\parallel}) / \delta \hat{P}_{\parallel} \gtrsim 1, \qquad (7)
$$

is a condition for the stochasticity called the Chirikov criterion, which determines when resonance overlapping occurs.

The main analysis of this study involves comparing simulation results with the prediction in Eq. (6) and describing the stochastic motion of electrons when condition (7) is satisfied. Solar wind electrons are typically non-relativistic ($\gamma \approx 1$). In that case, the H surfaces in the solar wind frame are described by $v_{\perp}^2 + (v_{\parallel} - \omega/k_{\parallel})^2 =$ const (circular contours centered around the wave phase velocity) and the resonant velocities are $v_{\parallel r} = (\omega - n\Omega_c)/k_{\parallel}$.

### B. Sensitivity to initial conditions

The derivations leading to (5) and (6) require certain approximations of the Hamiltonian $\mathscr{H}$. For simulating the full dynamics, the Lorentz equations

$$
\frac{d\mathbf{r}}{dt} = \mathbf{v}, \qquad (8a)
$$

$$
\frac{d(\gamma \mathbf{v})}{dt} = -\frac{e}{m} [\mathbf{E}_w + \mathbf{v} \times (\mathbf{B}_w + \mathbf{B}_0)] \qquad (8b)
$$

equivalently describe our system without such approximations. In Hamiltonian systems, the phase space volume, which is a function of energy, is conserved. Thus, the Boris method (Birdsall and Langdon, 1985), previously shown capable of preserving volume (Qin et al., 2013), is a natural algorithm for simulating the dynamics given by (8).

However, this volume-preserving characteristic is only well-maintained (over a long time) when the magnetic field is constant or the scalar potential is quadratic (Hairer and Lubich, 2018), neither of which is the case in our system where the fields (1) are periodic. Thus, the energy error might not be globally bounded (Hairer and Lubich,

2018; Zafar and Khan, 2021). In a small enough time period $\Delta t$, however, both the magnetic field and the potential can be approximately constant and quadratic, respectively, through a Taylor expansion. Therefore, it is necessary to determine the $\Delta t$ for which this occurs. In the following, we describe a method to achieve this through an estimation of the phase space volume. Consequently, this is also a measure of the efficiency of the Boris method at resolving the particle dynamics when the waves are high amplitude.

The dynamical system (8) can be written as $d\mathbf{X}/dt = \mathbf{F}(t, \mathbf{X})$ where $\mathbf{X} = (\mathbf{r}, \gamma\mathbf{v})$ is a unique particle trajectory in 6D phase space, given an initial condition $\mathbf{X}(0) = \mathbf{X}_0$. An arbitrarily small displacement $\boldsymbol{\delta}$ from $\mathbf{X}$ will evolve in time as dictated by the Jacobian $\nabla \mathbf{F}(t, \mathbf{X})$

$$
\frac{d\boldsymbol{\delta}}{dt} = \boldsymbol{\delta}^T \cdot \nabla \mathbf{F}, \qquad (9)
$$

where $\boldsymbol{\delta}^T$ is the transpose of $\boldsymbol{\delta}$. Stochastic motion is highly sensitive to initial conditions, meaning an initially small $\boldsymbol{\delta}$ may grow exponentially large. A measure for such stochasticity is the Lyapunov characteristic exponent (LCE), formally defined as the mean growth rate in $\boldsymbol{\delta}$ (Lichtenberg and Lieberman, 1992), given by

$$
h_{\delta} \equiv \lim_{\substack{t \to \infty \\ \delta \to 0}} \left(\frac{1}{t}\right) \ln \frac{||\boldsymbol{\delta}(t)||}{||\boldsymbol{\delta}(0)||}. \qquad (10)
$$

Since our phase space is 6D, there is a spectrum $\mathscr{S} = \{h_i\}_{i=1}^6$ of the LCE corresponding to the growth rate in the $i$th dimension of $\mathbf{X}$. Trajectories close to $\mathbf{X}$ will either diverge ($h_i > 0$), converge ($h_i < 0$), or remain at the same distance ($h_i = 0$) in each dimension with rates in time described by the LCE spectrum $\mathscr{S}$.

In describing the stochasticity, an important quantity is the maximal LCE, $\max(\mathscr{S})$, as has been used in the study of Wykes et al. (2001). However, for our study, we focus on the sum of the LCE spectrum, or the total LCE, $h \equiv \sum_{i=1}^6 h_i$. Conservation of the phase space volume requires that $h = 0$. If we define the (time-averaged) relative volume expansion as

$$
\frac{\Delta V(t)}{V_0} \equiv \exp(ht) - 1. \qquad (11)
$$

Then, $\Delta V/V_0 = 0$ whenever the volume is conserved. This condition regarding volume expansion is more physical than the growth rate $h$, since it describes a property of local groups of solutions in phase space.

In practice, it is possible to choose a $\Delta t$ such that $\Delta V/V_0$ is smaller than a reasonable threshold. Thus, we can ensure that the phase space volume around our solution is reasonably conserved, which implies from the study of Hairer and Lubich (2018) that the energy error is bounded. In Sec. III, we provide some demonstrations of this method through calculations of (11).

### III. SIMULATION METHODS

In this study, we use a *variational test particle simulation* to investigate the interactions of solar wind electrons with monochromatic whistler-mode waves. Realistically, whistlers are observed in a spectrum of frequencies and wave vectors ($\omega_0 \pm \Delta\omega, \mathbf{k}_0 \pm \Delta\mathbf{k}$). In that case, there is always resonance overlapping, independent of amplitude, at the same harmonic among waves in the spectrum. However, we wish to study the overlapping between different harmonics when the amplitude is high and the diffusion may be much more significant. Many solar wind whistlers are in fact narrowband (Cattell *et al.*, 2020; 2021b; Agapitov *et al.*, 2020), so this is reasonable in certain conditions.

An advantage of the test particle approach is the freedom to design the wave fields at the expense of not simulating the self-consistent evolution of the waves and particles, which might lead to unphysical effects in the particle motion. However, as have been compared in Cattell and Vo (2021), the VDF calculated from this approach shows no contradictory behaviors with those in the self-consistent simulation of Micera *et al.* (2020), which is hereby referred to as M2020. Thus, we can be assured in the context of that study that our solutions are consistent with PIC results. However, our approach is not only limited to test particle simulations. This will be further discussed in Sec. V.

We use the relativistic Boris algorithm (Ripperda *et al.*, 2018) to solve (8) numerically with a range of initial conditions. Similar to Cattell and Vo (2021), we study two sets of background parameters typical of solar wind conditions at 0.3 and 1 AU. The former, identical to those in M2020, has an electron density $n_e = 350 \, \text{cm}^{-3}$ and a background field $B_0 = 60$ nT so that $\omega_{pe}/\Omega_c = 100$ where $\omega_{pe} = \sqrt{e^2 n_e/\varepsilon_0 m}$ is the plasma frequency. The latter has $n_e = 5 \, \text{cm}^{-3}$, $B_0 = 10$ nT, and $\omega_{pe}/\Omega_c = 71$. The wave frequency is $\omega/\Omega_c = 0.15$, typical of observed solar wind whistlers (Cattell *et al.*, 2020). For comparison, this is about 1.5–3 times larger than that of the oblique whistlers in M2020. Although the frequency of quasi-parallel whistlers in M2020 was not reported, a simulation with similar parameters in Micera *et al.* (2021) with an expanding box model observed comparable frequency between oblique and quasi-parallel whistlers. Thus, we do not vary frequency in this study. It has minimal significance in our later discussions because the amount of overlap is mostly affected by amplitude. The effects of varying amplitude and propagation angle are studied in Sec. IV.

The variational aspect of the simulation comes from the LCE spectrum calculations, enabling the computation of (11). A variational method of estimating $\mathscr{S}$ for Hamiltonian flows has been demonstrated for a number of smooth dynamical systems (Benettin *et al.*, 1980; Sandri, 1996). It involves tracing the relative evolution of a tangent space along **X** under a local expansion operator

$$\mathbf{M}(t, \mathbf{X}) = \mathbb{1}_6 + \Delta t \nabla \mathbf{F}, \tag{12}$$

described by the dynamics (8), where $1_6$ is the 6D identity matrix. The Appendix provides a detailed discussion of the variational calculations of $h_i \in \mathscr{S}$. As a demonstration, a 2D example of such an evolution of the tangent space is given in Fig. 1(a). Under $n$ actions of **M** (or after a time period $n\Delta t$), the phase space around the particle might shrink or grow in certain dimensions. In this example, the rate is characterized by $h_1 > 0$ and $h_2 < 0$.

Figure 1(b) shows an example that is more relevant to our later simulations. The time evolution of a 6D LCE spectrum, typical of an electron interacting with a whistler in 1 AU parameters with $E_0 = |\mathbf{E}_w| = 1$ mV/m, $\delta B/B_0 = |\mathbf{B}_w/B_0| \sim 0.01$, and $\alpha = 65°$ is plotted in terms of the wave period $T_w = 2\pi/\omega$ using a time step $\Delta t/T_c = 10^{-5}$ where $T_c = 2\pi/\Omega_c$ is a cyclotron period. The particle has an initial kinetic energy $W = (\gamma - 1)mc^2 = 10$ eV and initial pitch angle $P = \cos^{-1}(v_z/v) = 0°$. Note that the formal definition

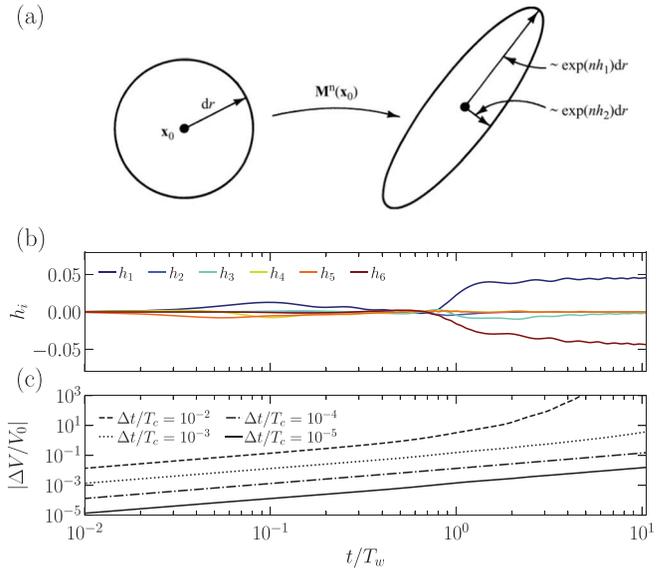

**FIG. 1.** Example of the variational calculations. (a) A visualization of the expansion of a 2D volume around the particle trajectory $\mathbf{x}_0$ after $n$ actions of **M**. Reproduced with permission from Ott, *Chaos in Dynamical Systems* (Cambridge University Press, 2002). Copyright 2002 Cambridge University Press [Ott (2002)]. The LCE $h_1$ and $h_2$ describe the exponential growth/decay along each principal axis of the volume. (b) The time evolution of a 6D LCE spectrum of a particle interacting with a whistler wave [stretches in the *x* and $p_z$ directions. (c) A comparison in the volume expansion among simulations with different $\Delta t$, as indicated in the legends. The early slope of all lines is close to 1, indicating a linear growth.

(10) of $h_i$ is a limit, so convergence must occur at large time periods. As expected for a periodic perturbation, this happens after one $T_w$. The convergent values also come in pairs $(h_i, h_j)$ where $h_i = -h_j$. In this case, the only non-zero pair is $(h_1, h_6)$ where $h_1 = \max(\mathscr{S}) = -\min(\mathscr{S})$, so the phase space evolves somewhat similarly to the sketch in panel (a) of Fig. 1. This symmetry naturally leads to $h = 0$ and is characteristic of Hamiltonian flows (Lichtenberg and Lieberman, 1992, p. 301).

Given $\mathscr{S}$, the volume expansion $\Delta V/V_0$ is calculated in Fig. 1(c). For comparison, those from a few similar simulations performed with different values of $\Delta t/T_c$ are also plotted. Initially, all of them exhibit a linear growth in time (the slope in the log–log plot is ~1), consistent with the reported behavior of the Boris algorithm (Hairer and Lubich, 2018; Zafar and Khan, 2021). However, at later times $t \geq T_w$, only simulations with $\Delta t/T_c \lesssim 10^{-4}$ have $\Delta V/V_0 \lesssim 0.1$, while that from simulations with larger time steps grows significantly large. Also, note that since $\Delta V/V_0$ usually grows monotonically, there is an implicit restriction on the maximum simulation run time for a given $\Delta t$ ($t_{max}/T_w < 10$ for the presented cases).

Having investigated the volume expansion around one particle trajectory, we now repeat the calculations for a set of electrons with $W$ from 0 to 3 keV and $P$ from 0 to 180°. Figure 2 shows $\Delta V/V_0$ around these particles after ~10 periods of interaction with an oblique ($\alpha = 65°$) whistler with (a) $\delta B/B_0 \sim 0.01$ and (b) $\delta B/B_0 \sim 0.1$. The overlaid lines corresponding to different harmonics $n$ show the resonance condition (5). In both panels, the resonance widths bounding "islands" around each harmonic are clearly observed. The islands are wider as the wave amplitude increases, and the volume expansion also becomes less uniform. Inside each island, electrons are trapped and have regular (quasi-periodic) motion with minimal $\Delta V/V_0$ (dark-blue regions). Outside, they are scattered and have more stochastic motion with larger $\Delta V/V_0$ (red regions). These red regions form a stochastic width around the resonant islands. Particles from these regions may eventually be trapped within a resonant island. At high amplitudes, the stochastic widths may overlap, resulting in island destruction or modification.

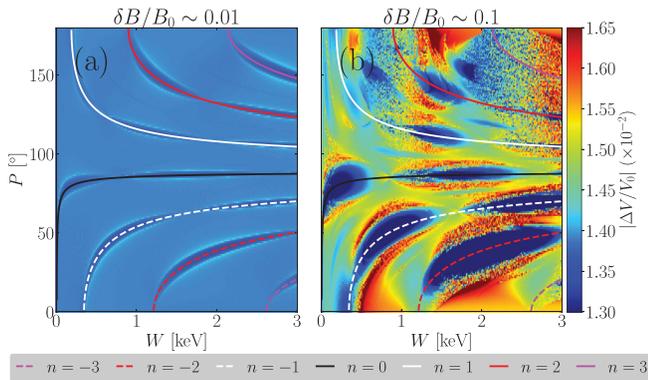

FIG. 2. Volume expansion around electrons with initial kinetic energy $W$ and pitch angle $P$. (a) $E_0 = 2$ mV/m, $\delta B/B_0 = 0.01$. (b) $E_0 = 20$ mV/m, $\delta B/B_0 = 0.1$. In both cases, $N_{\parallel} = 200$ (0.3 AU background parameters). The overlaid lines show the resonance condition for different harmonics as indicated in the legend. Solid lines ($n > 0$) correspond to the normal cyclotron resonances and dashed lines ($n < 0$) correspond to the anomalous cyclotron resonances.

As illustrated above, resonant interactions might lead to drastically different dynamics, resulting in non-uniform volume expansion among particles with different initial conditions. At high amplitudes, $\Delta V/V_0$ may increase quickly [as is the case in Fig. 1(c)] in some regions, while being minimal in others. Therefore, it is important to choose a $\Delta t$ such that there are on the same order of magnitude everywhere, ensuring consistency among all particle solutions. We find that a time step $\Delta t/T_c = 10^{-5}$ is a good choice to maintain $|\Delta V/V_0| \sim 10^{-2}$ (see color bar limits of Fig. 2). The same time step was used in the test particle simulations in Cattell and Vo (2021) and the PIC simulation in M2020. In Sec. IV, we study the stochastic motion of electrons in large amplitude waves using this time step.

## IV. RESULTS

In this section, we use the resonance-diagram technique (Karimabadi et al., 1990; Karimabadi et al., 1992) to study the 6D electron motion in phase space. To reveal the constants of motion, we only plot the surfaces of section (intersections of particle trajectories with $\dot{\theta} = \pi/2$ and $P_y = 0$). These intersections will trace out a continuous line in phase space if a constant of the motion (adiabatic invariant) is conserved (Lichtenberg and Lieberman, 1992, p. 48–52).

Figure 3 shows the intersections from the trajectories of electrons interacting with an oblique ($\alpha = 65°$) whistler in 0.3 AU (a)–(d) and 1 AU (e) parameters. Panels (a1)–(e1) of Fig. 3 show the surfaces of section in $(k_{\parallel}\hat{z}, \tilde{P}_{\parallel})$ phase space, where the wave phase $k_{\parallel}\hat{z} \sim k_{\parallel}z - \omega t$ and the parallel canonical momentum $\tilde{P}_{\parallel} = P_z$ are defined in Sec. II A. Panels (a2)–(e2) of Fig. 3 show those in the velocity ($v_z$, $v_{\perp}$) space. The underlying resonance islands located at a resonant velocity $v_{\parallel r}$ with effective width $\Delta v_{\parallel r}$ given in Eq. (6) are plotted as colored solid ($n > 0$) and dashed ($n < 0$) lines. Constant H surfaces going through $v_z = v_{\parallel r}$ and $v_{\perp} = 0$ are the black circular contours. The electrons are initialized with $v_{z0} = v_{\parallel r}$ corresponding to the $|n| = 0, 1, 2$ harmonics and $0 \leq v_{\perp 0} \lesssim 0.1c$ (in the middle of the islands).

Moving from (a) to (d) in Fig. 3, the whistler amplitude is increased from $\delta B/B_0 = 0.02$ to about 0.2 to show that the resonance islands gradually begin to overlap. In panel (a) of Fig. 3, the stochasticity condition [Eq. (7)] is not satisfied with $C \approx 0.3 < 1$, so the islands are well-separated and no overlapping occurs. The amplitude of the quasi-periodic motion of trapped electrons around $v_{\parallel r}$ agrees with the analytical prediction [Eq. (6)]. In panel (c) of Fig. 3 where $\delta B/B_0 \sim 0.1$, $C \approx 1.1$ and the islands at $|n| = 2$ (blue) and $|n| = 3$ (green) start overlapping. Traces of island destruction are seen in the top-most islands in panel (c1) of Fig. 3 for $0 \leq k_{\parallel}z - \omega t \leq \pi$ and in the bottom-most islands for the entire range of the wave phase. There is no longer any conserved quantity because the islands are destroyed. The electrons can then diffuse stochastically in phase space across multiple resonances. From panels (a2) to (d2) of Fig. 3, this diffusion is mostly constrained in the corresponding H surfaces, leading to a substantial change in the pitch angle as the resonance width increases.

The wave amplitude in panel (d) of Fig. 3 is one of the largest in observations (20 mV m$^{-1}$, although amplitudes >40 mV m$^{-1}$ are sometimes observed), and shows the same level of fluctuations in M2020. While the two bottom-most islands in panel (d1) of Fig. 3 (associated with the normal $n > 0$ cyclotron resonances) are significantly destroyed, the second ($n = -1$) and third ($n = 0$) islands from the top remain well-separated. In panel (d2) of Fig. 3, there is no diffusion of particles between them (dashed red and solid black islands). Similarly, there is minimal diffusion between the dashed red ($n = -1$) and dashed blue ($n = -2$) islands. Thus, while horn-like structures in the VDF form as strahl electrons are scattered along the H surfaces connected to these anomalous ($n < 0$) islands, they cannot be diffused to the Landau resonance $n = 0$ (black island) if the only interactions are with oblique whistlers. As shown in Cattell and Vo (2021), this is true even in the case of a packet of frequencies with the same bandwidth ($\sim$40–50 Hz) as those consistently generated from the PIC simulation in M2020.

Panels (d) and (e) of Fig. 3 have the same electric field magnitude (20 mV m$^{-1}$), which has been reported in 0.3 and 1 AU observations (Agapitov et al., 2020; Cattell et al., 2020; 2021b). However, since the background field far from the Sun is smaller, $\delta B/B_0$ is larger in Fig. 3(e) (around 0.7). In this case, the resonance overlap is so large ($C \approx 3$) that almost all electrons diffuse through the entire range of pitch angle ($0 \leq P \leq 85$). In panel (e1) of Fig. 3, only a few particles trapped in Landau resonance (moving at the wave phase velocity) still trace a continuous Poincare section bounded by the theoretical trapping width. However, these lines are modified compared to these cases in Figs. 3(a1)–3(d1), suggesting that these particles follow a different constant of motion. Overall, most particles are scattered isotropically, as seen in the 1 AU simulations in Cattell and Vo (2021).

Figure 4 investigates the effects of increasing propagation angle in high amplitude whistlers in 0.3 AU parameters. From Figs. 4(a)–4(d), $\alpha$ changes from 10° to 40°, while the amplitude is kept constant at 20 mV m$^{-1}$. To focus on the scattering between the anomalous ($n < 0$) resonances and the Landau resonance ($n = 0$), we only initiate electrons at the resonant velocities $v_{\parallel r}$ that correspond to $n = -1$ (dashed red) and $n = -2$ (dashed blue). For comparison with the scattering by high amplitude and oblique whistlers, we have included the $n = 0$ (black), $n = -1$ (red), and $n = -2$ (blue) resonance islands from Fig. 3(d2) as colored regions. The H contours connected to these islands are given the same colors to differentiate them from those associated with the simulated waves (black).

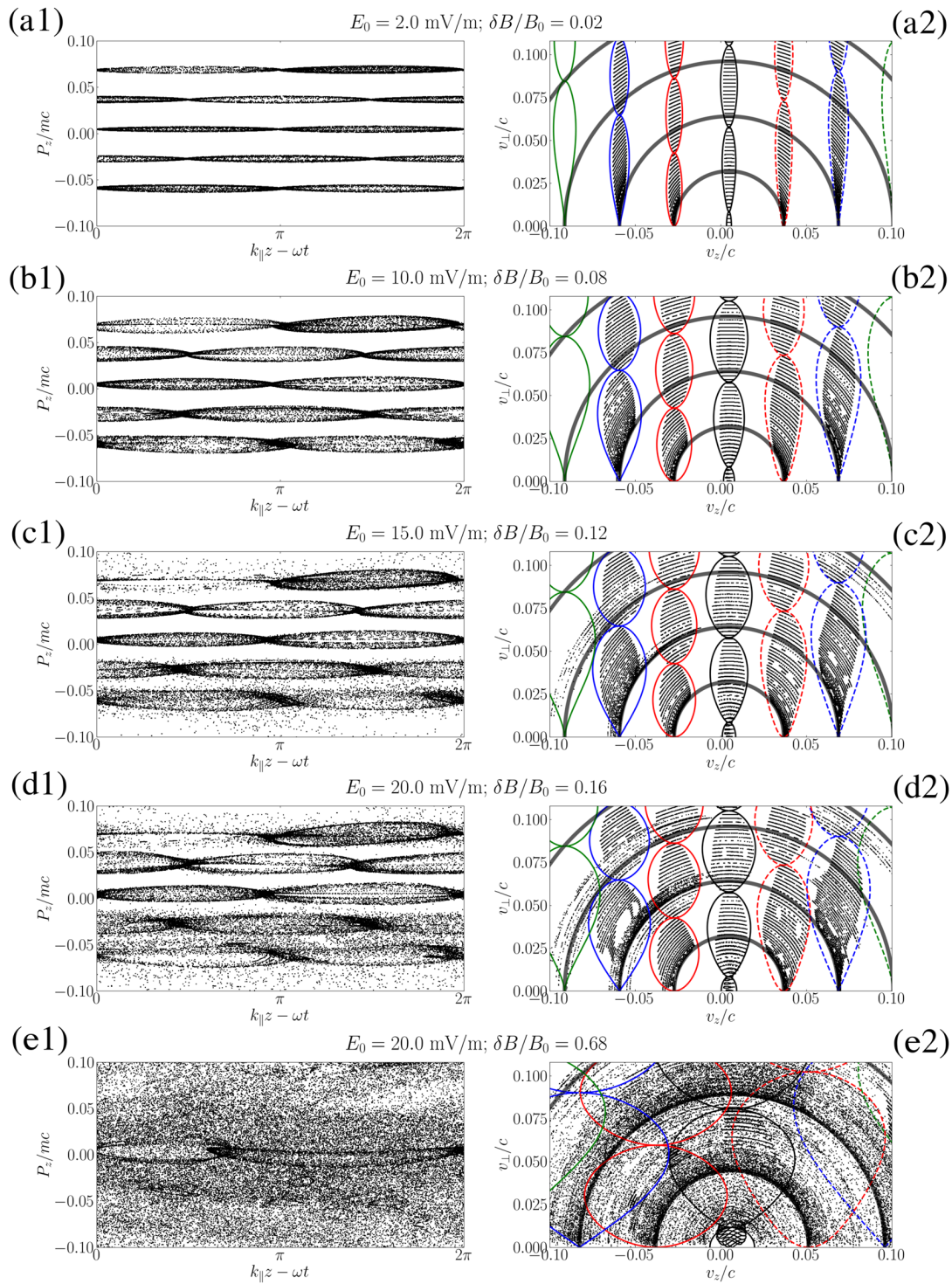

**FIG. 3.** Surfaces of section (at $\theta = \pi/2$ and $P_y = 0$) from the trajectories of electrons interacting with an oblique ($\alpha = 65°$) whistler in 0.3 AU [(a)–(d), in increasing amplitude] and 1 AU (e) parameters. (a1)–(e1) Intersections in the ($k_\parallel z, P_\parallel$) space and (a2)–(e2) intersections in the velocity ($v_z, v_\perp$) space. The colored lines are the resonance islands located at $v_z = v_{\parallel r}$ with width $\Delta v_{\parallel r}$ corresponding to $|n| = 3$ (green), $|n| = 2$ (blue), $|n| = 1$ (red), and $n = 0$ (black). Solid lines are the normal cyclotron resonances ($n \geq 0$). Dashed lines are the anomalous cyclotron resonances ($n < 0$). The electrons are initiated in the middle of the islands with $v_{z0} = v_{\parallel r}$ corresponding to $|n| = 0, 1, 2$ and $0 \leq v_{\perp 0} \lesssim 0.1c$. The underlying black circular contours are the H surfaces going through $v_z = v_{\parallel r}$ and $v_\perp = 0$ with centers at $v_z = \omega/k_\parallel$.

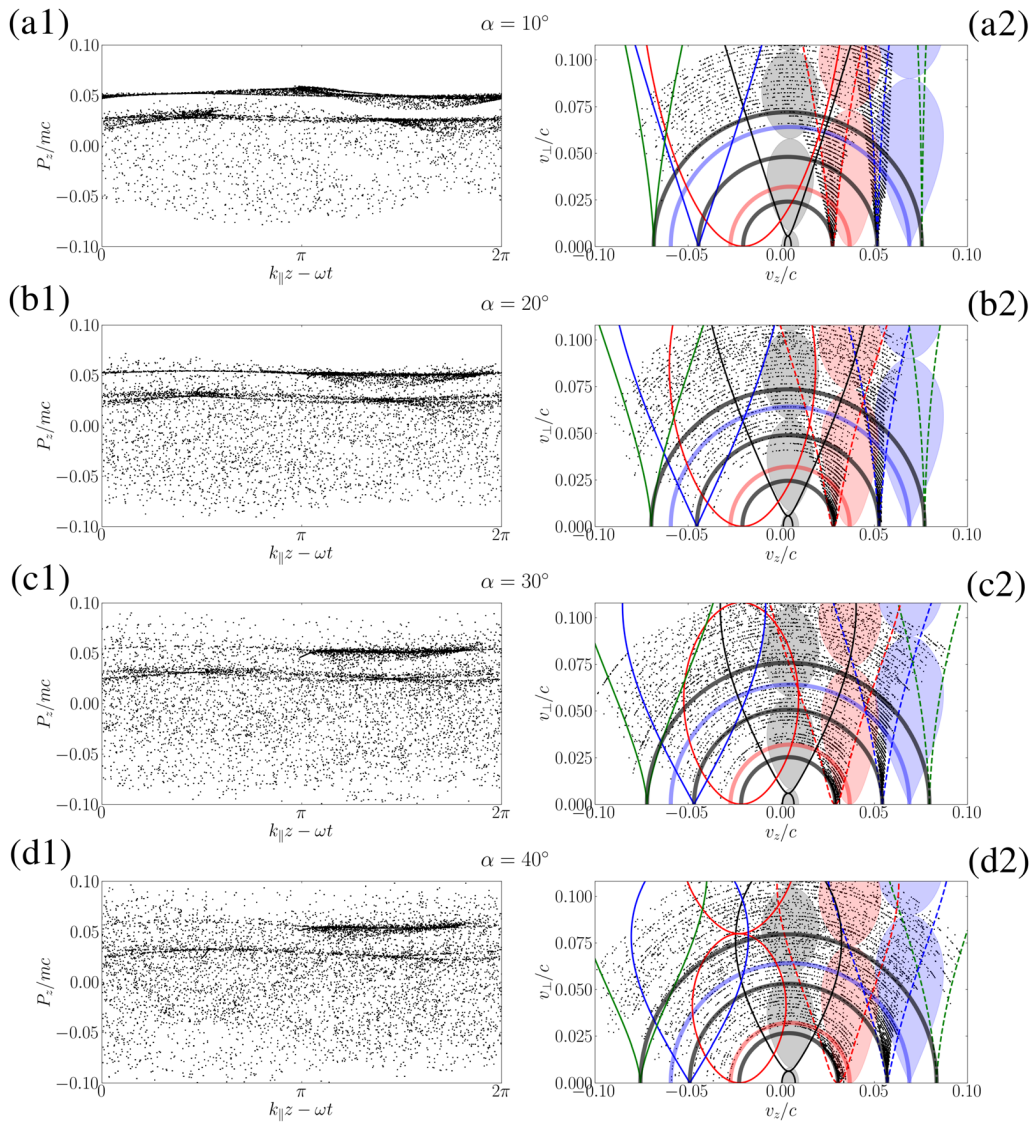

**FIG. 4.** Surfaces of section from trajectories of electrons interacting with a whistler in 0.3 AU parameters with $E_0 = 20$ mV/m, $\delta B/B_0 \sim 0.1$ and in increasing propagation angle $\alpha$ (a)–(d). The islands are plotted similar to those in Fig. 3. However, only electrons with $v_{z0} = v_{\parallel r}$ corresponding to $n = -1$ and $n = -2$ are initiated. The colored regions are the islands of 20 mV m$^{-1}$ oblique whistler in Fig. 3(d) corresponding to its $n = 0$ (black) $n = -1$ (red), and $n = -2$ (blue) harmonics. The H contours connected to these islands are also colored similarly, while those associated with the simulated waves are colored black.

In Figs. 4(a1)–4(d1), most particles have stochastic motion, except for those with low $v_\perp$. In Figs. 4(a2)–4(d2), electrons with high enough $v_\perp$ are significantly scattered through a wide velocity range ($-0.08c \lesssim v_z \lesssim 0.07c$, or almost the entire range of pitch angle) because of large resonance overlaps between different resonant harmonics. This means that although whistlers at low propagation angles are not effective in scattering highly field-aligned (strahl) electrons, they can isotropize a particle distribution as long as there is a mechanism that accelerates the electrons to a high enough $v_\perp$. For example, the waves in panels (a2) and (b2) of Fig. 4 can isotropize electrons around $v_z \approx 0.05c$ with $v_\perp \gtrsim 0.05c$. Strahl electrons (with predominantly parallel velocities) starting out in the blue region [$n = -2$ of oblique whistler in Fig. 3(d)] can continue to be scattered into the red region when they reach the dashed blue ($n = -2$) island along the blue circular contour.

At low wave angles, the fundamental cyclotron resonance ($n = 1$) has a very large resonance width, effective from $v_z \approx -0.07c$ to $v_z \approx 0.04c$ [see panels (a2) and (b2) of Fig. 4]. This $n = 1$ resonance also overlaps significantly with the Landau ($n = 0$) resonance. Thus, as strahl electrons are diffused close to $v_z \sim \omega/k_{\parallel} = 0.003c$, they quickly interact with both the $n = 0$ and $n = 1$ resonances and are further scattered to lower velocities ($v_z \lesssim \omega/k_{\parallel}$), resulting in an isotropic distribution function. In panels (c2) and (d2) of Fig. 4, the effective velocity range of the fundamental cyclotron resonance becomes smaller, as the whistler obliquity increases. However, the non-fundamental island widths also grow larger, resulting in more resonance overlaps, which could isotropize the distribution as demonstrated in Fig. 3.

## V. DISCUSSION AND CONCLUSIONS

The effects of increasing the whistler amplitude and propagation angle are studied through calculations of the resonance width. While high amplitude and oblique whistlers in 1 AU solar wind can form an isotropic halo, anomalous interactions with high amplitude and quasi-parallel whistlers may be essential to the process of halo formation at 0.3 AU. Our results indicate that quasi-parallel whistlers assume two roles. First, while we have shown that they cannot effectively scatter highly field-aligned strahl electrons, they can facilitate the pitch angle diffusion across different resonant harmonics of oblique whistlers in the range $v_z \geq \omega/k_{\parallel}$ because their resonant islands are located in the middle of those of oblique whistlers at high $v_{\perp}$ (see Fig. 4). Second, at high amplitudes, the effective velocity range of their fundamental harmonic is very large. Therefore, quasi-parallel whistlers can form an isotropic population after electrons are scattered to $v_z \sim \omega/k_{\parallel}$ due to the combined effects of quasi-parallel and oblique whistlers.

Whistlers observed by the Parker Solar Probe (PSP) in Encounter 1 generally propagate within $20°$ of the background field (Cattell et al., 2021b). The whistlers simulated in Figs. 4(a2) and 4(b2) are similar to those observed in Cattell et al. and quasi-parallel whistlers appearing at the later stages of M2020. For $\alpha \leq 40°$, they are all capable of isotropizing strahl electrons near the $n = -2$ resonance (blue contour), but only waves with $\alpha \geq 30°$ can scatter $n = -1$ electrons (along the red contour) to lower $v_z$. However, the overlap between the $n = 0$ and $n = -1$ harmonics of the $\alpha \leq 20°$ wave is very close to covering the energy range of the red contour. We speculate that a spectrum of waves should be able to isotropize these electrons, as is the likely case in M2020.

These conclusions provide insights into the particle diffusion observed in the PIC simulation results of M2020. The level of fluctuation ($\delta B/B_0 \approx 0.1$) and variation of propagation angle ($\alpha \sim 50°–70°$ for oblique whistlers and $\alpha \sim 0°–20°$ for quasi-parallel whistlers) used in our study are mostly consistent with the waves in their simulation. Without the presence of quasi-parallel whistlers, the strahl electron population at 0.3 AU is only scattered up to a certain degree and the resulting distribution is not isotropic (Cattell and Vo, 2021). However, since high amplitude and quasi-parallel whistlers exist at the same time that a completely isotropic halo is formed, they are likely to play a role in isotropizing strahl electrons as discussed above. Note that most previous discussions pertain to anti-sunward whistlers. In M2020, both sunward and anti-sunward quasi-parallel whistlers exist. The former mostly plays the same role as anti-sunward oblique whistlers when the wave amplitude is high because its fundamental cyclotron resonance is located at the strahl energy range and the corresponding resonant width is very large.

In the PIC simulation in Roberg-Clark et al. (2019) examining solar flares, the separation between consecutive islands is large, as the electron energy range is relativistic. Thus, the existence of electrostatic waves is necessary, as they play the same role in pitch angle diffusion among different harmonics as quasi-parallel whistlers do in our simulation. In the solar wind, the bulk of the electrons are at non-relativistic energies and the islands are more closely separated. Thus, when the amplitude is high and when there is a wide enough spectrum, the processes of strahl scattering and halo formation can be completely carried out by whistler-mode waves.

At 1 AU, observations of quasi-parallel whistlers (Lacombe et al., 2014; Tong et al., 2019b) show that they are usually low amplitude ($\delta B/B_0 \lesssim 0.01$). However, since our results indicate that large amplitude, oblique whistlers alone can isotropize the distribution [see Fig. 3(e)], the process of strahl scattering at 1 AU is likely facilitated by them. Near the Sun, solar wind observations ($\lesssim 0.3$ AU) show that high amplitude whistlers, both quasi-parallel and oblique, exist (Agapitov et al., 2020; Cattell et al., 2021b). Thus, quasi-parallel whistlers may play a more important role in strahl scattering and halo formation at shorter heliospheric distances.

Breneman (2010) and Cattell et al. (2020) provided statistics on the occurrence of high amplitude and oblique whistlers observed by STEREO at 1 AU, but did not report the relationship between amplitude and propagation angle, which are two important properties for the understanding of strahl scattering and halo formation. Figure 5 plots the whistler amplitude dependence on wave angle, determined from a database of whistler waveforms obtained by the STEREO S/WAVES waveform capture instrument (Bougeret et al., 2008) and described in Cattell et al. (2020). The high amplitude whistlers observed at 1 AU tend to be highly oblique, with most propagation angles ranging from $45°$ to $70°$, near the resonance cone angle. $\delta B/B_0$ is frequently in the range of 0.5–0.8. Thus, the typical interactions between 1 AU solar wind electrons (especially the strahl) with these waves are expected to be similar to the case presented in Fig. 3(e). While this provides some observational support for our claim at 1 AU, there has been no statistical study of the dependence of amplitude on wave angle near the Sun. Determining this property for whistlers near 0.3 AU is necessary to verify our conclusions.

The process of strahl scattering by oblique whistlers has also been studied with simulation models involving quasi-linear theory (Jeong et al., 2020). In their model, the wave amplitude is small ($\delta B/B_0 \sim 0.001$). Thus, the most scattering is achieved through resonance overlapping between the same harmonic of waves in a spectrum. However, at high amplitudes, overlapping between different harmonics results in a much wider range of pitch angle diffusion, as demonstrated through the resonance width calculations in Figs. 3 and 4. When there is a spectrum, the effective velocity range due to both types of resonance overlap becomes even larger. Thus, models based on a quasi-linear approach might need to adjust the effective width around each resonant harmonic when wave amplitudes are high. It is likely, however, that the diffusion in this case is non-linear and incompatible with the quasi-linear theory. Details of the nature of the resulting diffusion by high amplitude wave, however, are outside the scope of this paper.

In this study, we also presented a careful treatment of the sensitivity to initial conditions in numerical solutions of particle trajectories in the presence of high amplitude waves. The dynamics might be drastically different among a given range of initial conditions due to the highly stochastic motion through resonant interactions. Therefore, it is important to ensure consistency in terms of phase space volume conservation when considering the scattering of a particle distribution.

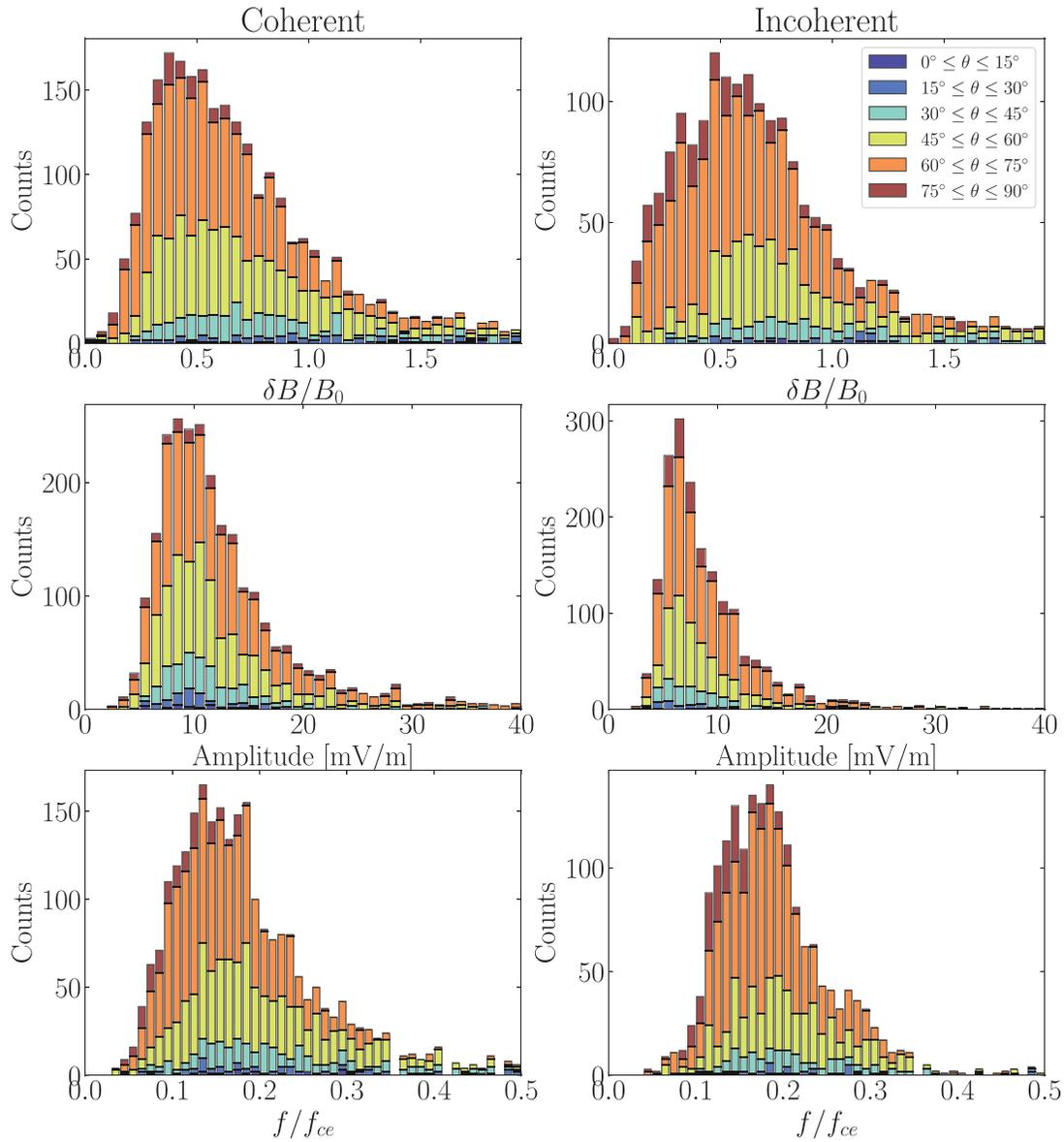

**FIG. 5.** Histograms of the relative amplitude $\delta B/B_0$ (top row), absolute amplitude in mV m$^{-1}$ (middle row), and relative frequency $f/f_{ce}$ (bottom row) of large amplitude whistler-mode waves observed by STEREO from March 2017 to September 2018. $f_{ce}$ is the cyclotron frequency. The histograms are color coded by propagation angles. Observed waves are categorized into coherent or incoherent waves based on the bandwidth.

The variational calculations that we described here can also be applied to PIC simulations because the particle advancing algorithm is the same. Thus, PIC studies might be conducted together with our method of determining the step size $\Delta t$ to ensure a homogenous behavior in the volume expansion $\Delta V$. That would enable investigations of more realistically evolved wave profiles, where there is no longer the lack of self-consistency. The calculation of resonance width may be generalized for a spectrum of wave, and the effects of a spectrum on the particle distribution may be better studied. However, that will be left for future studies.

## ACKNOWLEDGMENTS

The authors thank A. Artemyev, V. Roytershteyn, O. V. Agapitov, A. Micera, and G. T. Roberg-Clark for helpful discussions. They would also like to thank the reviewers for their suggestions that greatly improved the manuscript. The Minnesota

Supercomputing Institute (MSI) at the University of Minnesota provided the resources that contributed to the research results reported within this paper. URL: https://www.msi.umn.edu. This work was supported by NASA Grant Nos. NNX16AF80G, 80NSSC19K305, and NNN10AA08T, and NSF Grant No. AGS-1840891.

## AUTHOR DECLARATIONS

### Conflict of Interest

We have no conflict of interest to disclose.

### DATA AVAILABILITY

The data that support the findings of this study are available from the corresponding author upon reasonable request.

## APPENDIX: CALCULATIONS OF THE LYAPUNOV CHARACTERISTIC EXPONENTS

First, consider a basis of orthonormal vectors $\{\boldsymbol{\mu}_i\}_{i=1}^{6}$ forming a 6D parallelepiped $U$. The vectors $\boldsymbol{\mu}_i$ span the tangent space of a particle trajectory $\mathbf{X}$, where $\mathbf{X}(t)$ is a solution of the dynamical system (8). Using the wedge product, we can write $U = \boldsymbol{\mu}_1 \wedge \cdots \wedge \boldsymbol{\mu}_6$ and its volume

$$V(U) = ||\boldsymbol{\mu}_1 \wedge \cdots \wedge \boldsymbol{\mu}_6|| = \prod_{i=1}^{6} ||\boldsymbol{\mu}_i||. \tag{A1}$$

By assumption, the original volume $V_0 = 1$.

Following the evolution of $\boldsymbol{\mu}_i$ along $\mathbf{X}$ after a small time step $\Delta t$, we discretize (9) and substitute $\boldsymbol{\delta} = \boldsymbol{\mu}_i$. It follows that (up to first order in $\Delta t$)

$$\boldsymbol{\mu}_i' \equiv \boldsymbol{\mu}_i(t + \Delta t) \approx \mathbf{M} \cdot \boldsymbol{\mu}_i(t), \tag{A2}$$

where $\mathbf{M}(t, \mathbf{X}) = 1_6 + \Delta t \nabla \mathbf{F}(t, \mathbf{X})$ is an operator describing the evolution of the tangent space of $\mathbf{X}$ after a period $\Delta t$, $1_6$ is the 6D identity matrix, and $\boldsymbol{\mu}_i'$ are the deformed vectors after one action of $\mathbf{M}$. In the non-relativistic regime, the Jacobian $\nabla \mathbf{F}$ is

$$\nabla \mathbf{F} = \begin{pmatrix} 0 & \mathbb{1}_3 \\ (-e/m)D_r & (-e/m)D_v \end{pmatrix}, \tag{A3}$$

where $D_r = \nabla_\mathbf{r}(\mathbf{E}_w + \mathbf{v} \times \mathbf{B}_w)$ and

$$D_v = \nabla_\mathbf{v}(\mathbf{v} \times \mathbf{B}) = \begin{pmatrix} 0 & B_z & -B_y \\ -B_z & 0 & B_x \\ B_y & -B_x & 0 \end{pmatrix}. \tag{A4}$$

Since $\boldsymbol{\mu}_i$ might not be eigenvectors of $\mathbf{M}$, it is not guaranteed that $\boldsymbol{\mu}_i'$ form an orthogonal set. Thus, to calculate the volume of the space spanned by $\boldsymbol{\mu}_i'$, we can use the Gram–Schmidt orthogonalization procedure on $\boldsymbol{\mu}_i'$ to find a set of orthogonal vectors $\{\boldsymbol{\omega}_i\}_{i=1}^{6}$. Then, $V(U') = V(\boldsymbol{\omega}_1 \wedge \cdots \wedge \boldsymbol{\omega}_6)$ is the relative volume change of the original parallelepiped $U$. From (10), after a time $t_N = N\Delta t$, or $N$ actions of $\mathbf{M}$, we can approximate

$$h \approx \frac{1}{N\Delta t} \sum_{n=1}^{N} \ln \text{Vol}(U_n') = \frac{1}{N\Delta t} \sum_{n=1}^{N} \sum_{i=1}^{6} \ln ||\boldsymbol{\omega}_i^n||, \tag{A5}$$

where $U_n' = U'(t_n)$ and $\boldsymbol{\omega}_i^n = \boldsymbol{\omega}_i(t_n)$. Moreover, the LCE spectrum components are

$$h_i \equiv \frac{1}{N\Delta t} \sum_{n=1}^{N} \ln ||\boldsymbol{\omega}_i^n||, \tag{A6}$$

such that $h = \sum_{i=1}^{6} h_i$. Note that the orthogonal vectors $\boldsymbol{\omega}_i^n$ need to be normalized to unity after each time step so that the computed change is relative. Thus, (9) is solved variationally.